\documentclass[9pt,twocolumn,twoside]{osajnl}

\journal{ol} % Choose journal  (ao,jocn,josaa,josab,ol,optica,pr)

\setboolean{shortarticle}{true}
\usepackage{xcolor}
\usepackage{graphicx}% Include figure files

\title{Single photon emission from lithographically-positioned engineered nanodiamonds for cryogenic applications}
\author[1,$\dag$,*]{Vivekanand Tiwari}
\author[1,2,$\dag$]{Zhaojin Liu}
\author[1]{Hao-Cheng Weng}
\author[1]{Krishna C Balram}
\author[1]{John G Rarity}
\author[3]{Soumen Mandal}
\author[3]{Oliver A Williams}
\author[4]{Gavin W Morley}
\author[1,5,*]{Joe A Smith}

\affil[1]{Quantum Engineering Technology Labs, H. H. Wills Physics Laboratory and School of Electrical, Electronic, and Mechanical Engineering, University of Bristol, BS8 1FD, UK}
\affil[2]{Quantum Engineering Centre for Doctoral Training, H. H. Wills Physics Laboratory and School of Electrical, Electronic, and Mechanical Engineering, University of Bristol, BS8 1FD, UK}
\affil[3]{School of Physics and Astronomy, Cardiff University, Queen’s Building, The Parade, Cardiff CF24 3AA, United Kingdom}
\affil[4]{Department of Physics, University of Warwick, Coventry CV4 7AL, United Kingdom}
\affil[5]{School of Electrical and Electronic Engineering, University of Sheffield, Sheffield S1 3JD, United Kingdom}
\affil[$\dag$]{The authors contributed equally to this work.}
\affil[*]{vivekanand.tiwari@bristol.ac.uk, joe.a.smith@sheffield.ac.uk}
\begin{abstract} 
Nitrogen-vacancy centres in nanodiamonds (NDs) provide a promising resource for quantum photonic systems. However, developing a technology beyond proof-of-principle physics requires optimally engineering its component parts. 
In this work, we present a hybrid materials platform by photolithographically positioning ball-milled isotopically-enriched NDs on broadband metal reflectors. The structure enhances the photonic collection efficiency, enabling cryogenic characterisation despite the limited numerical aperture imposed by our cryostat. Our device, with SiO$_2$ above a silver reflector, allows us to perform spectroscopic characterisation at 16 K and measure autocorrelation functions confirming single-photon emission (g$^2$(0)<0.5). Through comparative studies of similar hybrid device configurations, we can move towards optimally engineered techniques for building and analysing quantum emitters in wafer-scale photonic environments.   
\end{abstract}

\setboolean{displaycopyright}{false}

\begin{document}
\maketitle
%%%%%%%%%%%%%%%%%%%%%%%%%%  body  %%%%%%%%%%%%%%%%%%%%%%%%%%
%\section{Introduction}
For quantum information processing applications, the nitrogen-vacancy (NV) centre defect in nanoscale-inclusions of diamond, termed nanodiamond (or ND), offers a lithographic means to combine spin-photon interfaces with both foundry photonics and microelectronics at scale \cite{weng2023heterogeneous,weng2024crosstalk}. Here, ND lithographic deposition overcomes a significant limitation: localisation of a stochastically formed atom-like emitter.

However, a key barrier to adoption of ND-based spin-photon interfaces is the proximity of the diamond surfaces to the atom-like system, which typically introduces unwanted charge noise and strain. Significantly, especially for the NV centre, this was enough to consider the platform not suitable for quantum information applications \cite{faraon2011resonant}. Recently, results have begun to challenge this assumption, with high optical and spin coherence reproducibly reported in ND that has been engineered specifically for quantum purposes \cite{orphal2023optically,march2023long}. This is a significant technology development compared to earlier results which used ND powders synthesised primarily as a mechanical lubricant \cite{knowles2014observing,smith2020single}. The prospect of combining ND with a chip-scale platform lends itself to then adopting semiconductor techniques such as encapsulation \cite{smith2020single} and surface engineering \cite{kumar2024stability}, which may in turn lead to the development of a technologically-mature platform for quantum information processing, especially compared to bulk diamond, which is costly, limited in substrate size, and difficult to integrate both photonic structures and active electronics. 

For photonics, significant efforts have been made to address the limitations caused by isotropic angular emission patterns and low emission rates in high-index semiconductor-hosted quantum emitters by modifying their surrounding photonic environment \cite{lubotzky2025approaching, wan2018efficient}. For example, the use of a circular Bragg resonator for a solid-state source of strongly entangled photon pairs with high brightness and indistinguishability \cite{wang2019demand}, plasmonic bullseye antennas \cite{andersen2018hybrid, abudayyeh2017quantum} and deterministic enhancement of coherent photon generation using distributed Bragg reflectors \cite{riedel2017deterministic}. In general hybrid, or hetereogeneous, material platforms lend themselves well to creating quantum photonic interfaces as the photonic material can be selected for its quality and then independently combined with an optimised quantum component in a post-fabrication step \cite{moody20222022,weng2023heterogeneous}. 

One simple photonic device is the use of a metallic reflector \cite{ma2014efficient}. The reflector is inherently broadband, so requires no tuning between the emission wavelength of the quantum emitter and the photonic device used to enhance light extraction compared to resonant structures \cite{haws2022broadband}. In this letter, we present a new photolithographic method to pattern arrays of engineered ballmilled ND \cite{march2023long}, using a direct laser writer and combine them with an engineered photonic device. As a proof of concept, we illustrate the use of a broadband metallic reflector. As opposed to positioning techniques that rely on electron beam lithography \cite{heffernan2017nanodiamond}, the photolithographic process presented in this manuscript provides a route to rapid, low-cost wafer-scale deposition of localised emitters, useful to rapidly characterise the compatibility of engineered quantum emitters with a range of different engineered photonics platforms. %Furthermore, unlike suspended nanostructures such as photonic crystals \cite{uppu2020scalable} or nanobeam waveguides \cite{makhonin2014waveguide, reithmaier2015chip}, which are efficient but suffer from poor scalability due to mechanical fragility, the proposed design is built on a solid-state base, specifically SiO$_2$ buffer layers with fewer surface states at an oxide interface. This provides a more robust platform that can support large-scale implementation \cite{chen2022scalable}. 

%\Electron-beam lithography (EBL) provides superior resolution (down to sub-10 nm), which is critical for fabricating nanoscale features with high precision. However, EBL suffers from limited throughput due to its serial writing nature, making it less practical for large-area patterning. In contrast, photolithography, while limited in resolution (~100 nm range depending on the wavelength and optics used), offers vastly higher throughput due to its parallel processing capability, and is more amenable to wafer-scale fabrication, making it highly scalable for industrial applications.

Typically, ND characterisation at room temperature is carried out in a confocal microscope with a large numerical aperture (NA =0.95). Limited by the physical constraints of our cryostat, we require a large working distance from the sample plane, that then limits the available numerical aperture (NA = 0.7) of the objective lens (working distance 6.0 mm MY100X-806 Mitutoyo). We found that the reduced NA resulted in low count rates (2 kc/s per channel), precluding autocorrelation measurement of ND on standard substrates (Si/SiO$_2$) and similar characterisation. Faced by this obstacle, we were motivated to design a metal reflector positioned beneath the ND, drawing on our previous work in encapsulated metallic photonics for visible photonics \cite{smith2022toward}. The device increases the broadband collection efficiency of emitters under study, and we show this enables cryogenic autocorrelation measurement. The device also serves as a simple test bed towards the more complex multi-material structures enabled by our photolithographic ND positioning technique.

For the device to achieve constructive interference between the light emitted by the NV centre in the top direction and the reflected light by the metal mirror, the thickness of SiO$_2$ needs to be precisely engineered. The calculation is performed using Finite-Difference Time-Domain (FDTD) method in Lumerical where we use an electric dipole with the emission wavelength 637 nm to simulate the single-photon source, at the zero phonon line (ZPL) of the NV$^-$ centre \cite{jeske2017stimulated}.

\begin{figure}[htbp] 
 \centering\includegraphics[width=\linewidth]{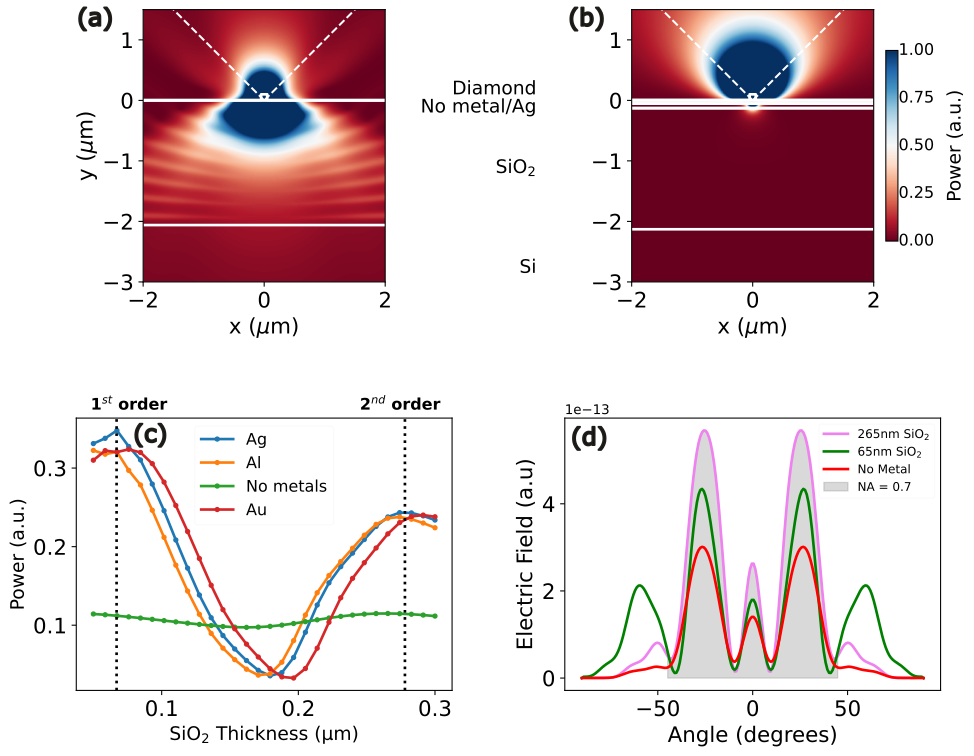}
 \caption{\label{fig1}   (a) Colour map showing power from a simulated NV dipole in diamond on a SiO$_2$ substrate, and (b) with the addition of a buried Ag metal reflector, dotted white line represents collection into Numerical Aperture NA=0.7. (c) Calculated emission power from NV dipole as a function of overgrown SiO$_2$ thickness above Ag, Al, Au and no metal reflector. Overgrown thicknesses corresponding to peaks emission from 1st order and 2nd order constructive interference are labelled. (d) Calculated electric field for metal (with 65 nm and 265 nm thickness of SiO$_2$) and no metal as a function of angle, with emission accepted by the NA=0.7 objective shown. 
 }
\end{figure}

In Figs. \ref{fig1} (a) and (b), the power flow without and with a metal reflector are illustrated, wherein a layer of silver (Ag) has been inserted into the material stack. Notably, without the metal reflector (Fig. \ref{fig1} (a)) a significant fraction of the power is lost in the downwards direction not accessible by the confocal microscope. 

Following on from this, in Fig. \ref{fig1} (c), we plot the collected power from a representative NV$^-$ dipole as a function of the buffer SiO$_2$ thickness for a range of metallic materials. The maximum power collection occurs around 65 nm of SiO$_2$, and a second order peak for all metals at around 265 nm  at the point of constructive interference, recording a maximum of three times enhancement compared to the absence of a metallic reflector. Fig. \ref{fig1} (c) shows superior reflectance using silver (Ag) to the other metals such as gold (Au) or aluminium (Al) \cite{osorio2012three}. Also, Ag metal reflectors are generally superior to Au for visible-light applications \cite{mcpeak2015plasmonic}. Ag exhibits a stronger electric field intensity and less light absorption as a result of its smaller imaginary dielectric component. This leads to significantly greater photoluminescence (PL) enhancement (e.g. 4.0-fold compared to Au's 1.8-fold) \cite{oto2021comparison}.  Experimentally, we have observed that Au metal reflectors can increase background noise, specifically originating from the interface between the Au film and the SiO$_2$ spacer \cite{monroy2022spatial}. In addition, Ag is both more reflective and more stable than Al.  %Ag  offers a wide controllable wavelength range, especially for blue/near-UV emissions, providing better energy matching \cite{fan2020improving, okamoto2004surface}.

With the optimal thickness chosen, it will be useful to calculate the far-field distribution to find the fraction accepted within the NA = 0.7 objective (45$^\circ$ half-angle). Fig. \ref{fig1} (d) shows the total power detected in the top plane. We take a centre line slice (Y-parallel in this case) of the far field hemispherical cone to generate a two-dimensional plot, as shown in Fig. \ref{fig1} (d). The shaded region corresponds to NA = 0.7, accepted emission. The percentage within the 0.7 NA is $5.98\%$, $11.5\%$ and $13.6\%$ for the device without a metal reflector, 65 nm and 265 nm SiO$_2$ respectively. This result is consistent with the near field power calculation, with the presence of the metallic mirror doubling the power collected. Detailed calculation methods can be found in Supplementary Information S2. Furthermore, the metallic structure enables efficient reflection over a broadband spectral range (Supplementary Information S2, Fig. S4). 

\begin{figure}[htbp]
 \centering\includegraphics[width=\linewidth]{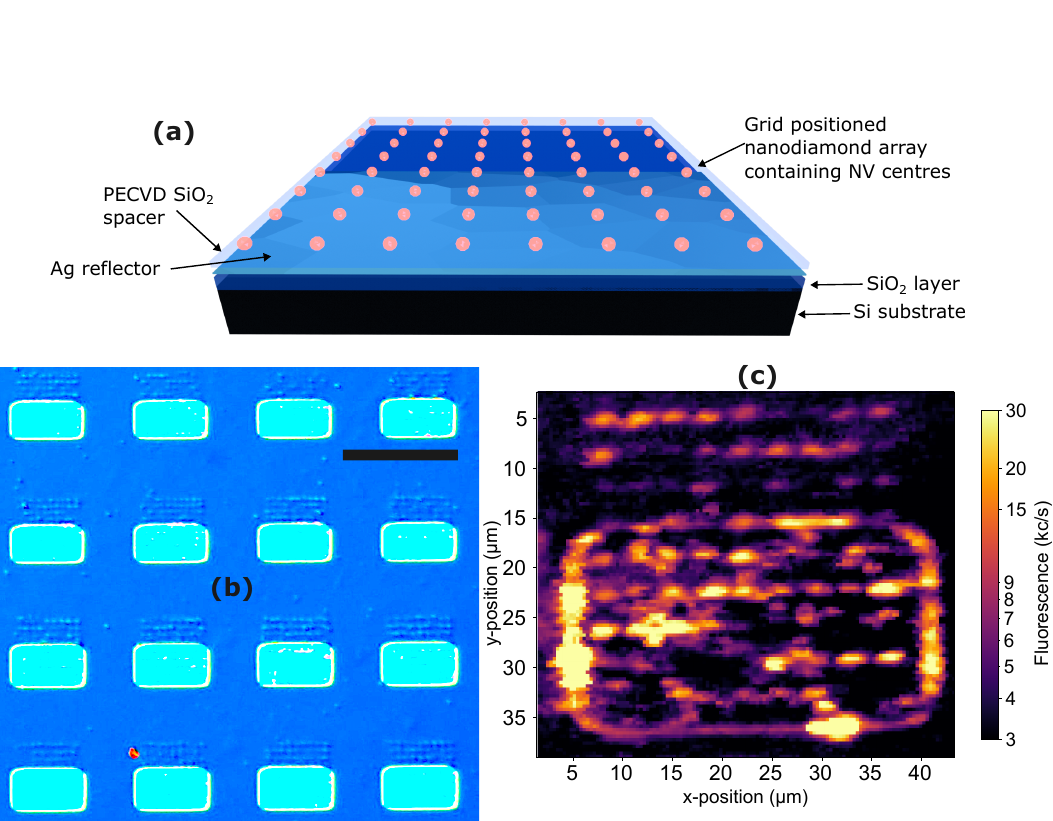}
\caption{\label{fig2} (a) Schematic presentation of the device with Ag metal reflector below the grown SiO$_2$ layer.  (b) Top view optical microscopic image of the photonic device showing the photolithographically positioned ND sites in an array (50 $\mu$m scale bar).  (c) Confocal map at 16 K over the metal and non-metal part of the device, indicating the presence of brighter NV centres correlated with the array sites on the metallic surface.
 }
\end{figure}

Following the design, we fabricated the hybrid photonic device including the photolithographically positioned NDs. First, the metal is deposited on a sample diced from a 2 $\mu$m thickness SiO$_2$ coated Si wafer. Then, a buffer layer of SiO$_2$ was grown by plasma-enhanced deposition (PECVD) with deposition rate of $75\pm 3$ nm/min  before isotopically-enriched ballmilled NDs are positioned \cite{march2023long}. Fig. \ref{fig2} (a), presents a schematic of the device, with half of the NDs deposited on top of the metal reflector and half on the dielectric for comparison. It was necessary to position the NDs using a bilayer resist with the top layer operating as a photoresist, and an underlying resist through which the pattern is transferred chosen for optimal selectivity in resist removal to ND adhesion. The complete fabrication process is discussed in Supplementary Information S3. Fig. \ref{fig2}(b) presents an optical microscopic image showing multiple successful copies of the photolithographically positioned ND array aligned to the metallic reflectors. A cryogenic confocal PL scan (16 K) of the device with 65 nm SiO$_2$ is presented in Fig. \ref{fig2} (c), where the metal boundary can be seen and bright Gaussian spots indicate the presence of NV centres in many of the positioned NDs. Of note, brighter PL (approximately 30 kc/s) from positioned sites are seen on the reflector compared to the region off the reflector (10 kc/s) indicating the photonic device is enhancing emission. 

Following the indicative PL confocal scans, we performed a set of measurements on two representative bright spots located on devices with 65 nm and 265 nm SiO$_2$ respectively to benchmark the NV centers inside the positioned ND. Details of the experimental setup and measurement procedures are presented in Supplementary Information S5. 

\begin{figure}[htbp] 
 \centering\includegraphics[width=\linewidth]{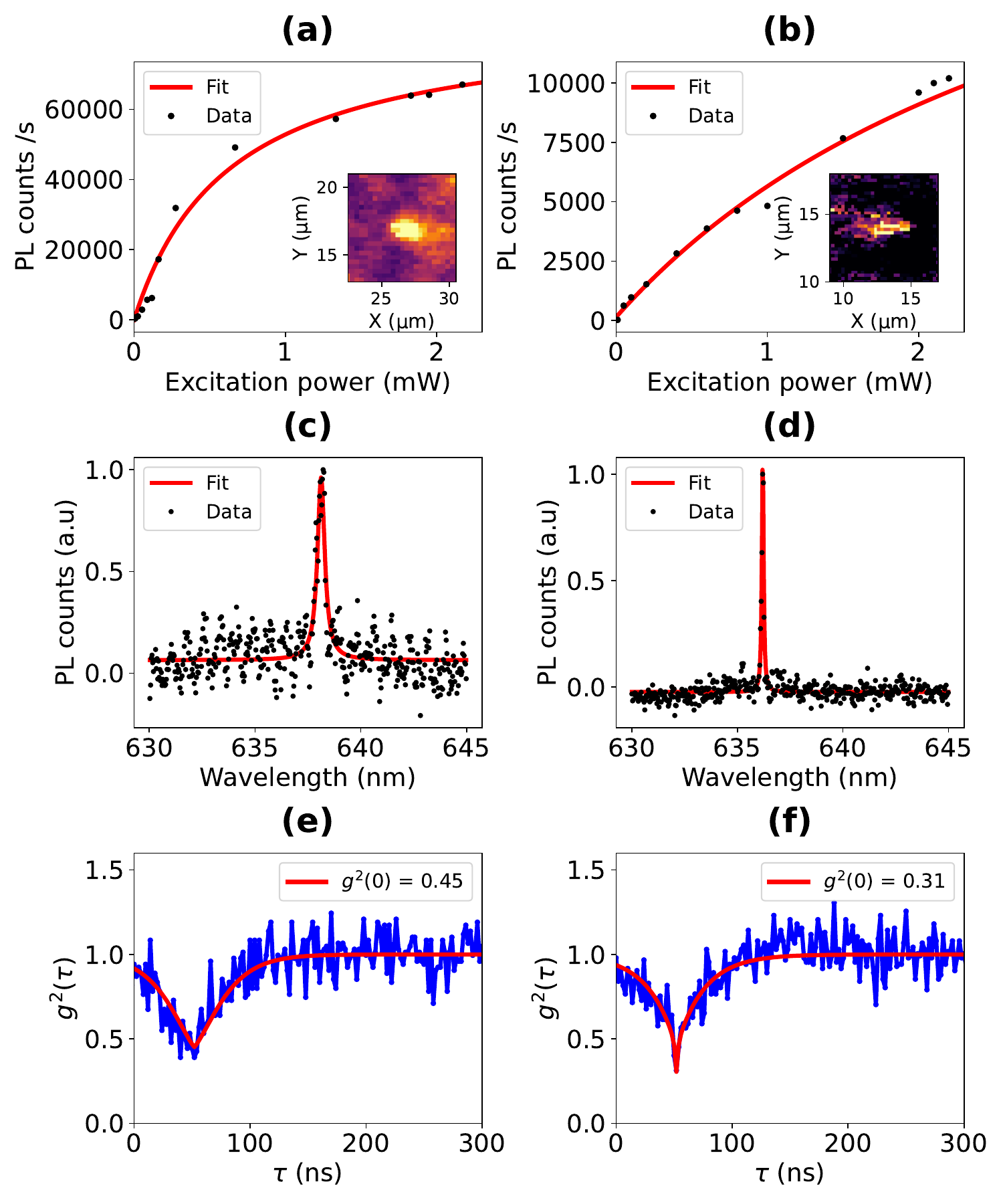}
 \caption{\label{fig3} Saturation power measurement at 16K of NVs in ND above Ag metallic reflectors with spacer layers of 65 nm SiO$_2$ (a) and 265 nm SiO$_2$ (b) with corresponding confocal scans inset. PL spectra at 16K of the NVs with 65 nm SiO$_2$ (c) and 265 nm SiO$_2$ (d) spacer layers. Intensity autocorrelation, g$^2$($\tau$) measurement of the NVs device with  65 nm SiO$_2$ (e) and 265 nm SiO$_2$ (f) spacer layers.
 }
\end{figure}

First, we take a saturation measurement by linearly increasing the laser excitation source and recording the PL. The plotted data is shown in Fig. \ref{fig3} (a) and (b). The background fluorescence contribution increases linearly with intensity, with single-photon fluorescence saturating on each ND site, modelled by the following expression: 
\begin{equation}
    I(P) = \frac{R P}{P_{\text{Sat}} + P} + nP + m
    \label{eq:example_equation}
\end{equation}
where $I(P)$ is the emission rate as a function of the laser power, $P$, saturating at $R$ with power $P_\text{Sat}$. In this expression, $n$ models a linear background and $m$ models the dark counts of the detectors \cite{weng2023heterogeneous}. The saturation power for the 65 nm SiO$_2$ device is at $P$ = 1 mW with $R$ = 86 kcount/s. In the case of a device with SiO$_2$ thickness of 265 nm the saturation power required is four times larger (4 mW) than the 65 nm SiO$_2$ thickness device with $R$ = 24 kcount/s. The trend correlates with our model, owing to
the modification of the local density of states  \cite{anger2006enhancement} from the reflector structure and enhancement in the emission from the NVs closer to metal surface \cite{choy2011enhanced, schietinger2009plasmon}. However, the excess brightness of the 65 nm layer sample may be due to noise from the close metallic interface or the presence of multiple emitters in the deposited ND. The inset confocal data shows that the background near to the NV is much brighter for the 65 nm device which is thought to be due to the closer metallic layer. 

Figs \ref{fig3} (c) and (d) show 16 K PL spectra for the two emitters. Both display a clear Zero Phonon Line (ZPL) at the known emission line for the NV centre, fitted using a single Lorentzian peak. The 265 nm sample linewidth has a narrow FWHM of 65 GHz, with this measurement ultimately limited by the resolution limit of the spectrometer (30 GHz). In contrast, the 65 nm line is around five times broader, which may be due to either the presence of multiple emitters or charge noise from the closer Ag surface. Additional spectra and fitting details are presented in Supplementary Information S6. More accurate linewidths could be measured by resonant PL excitation \cite{fu2009observation}. We do not expect these NV centres in ND's to be Fourier linewidth limited. However, placing photonic nanostructures around them could speed up emission via a strong Purcell factor, broadening the homogeneous linewidth and bringing emitted photons into the domain of indistinguishability \cite{smith2021nitrogen}.

Figs. \ref{fig3} (e) and (f) present the autocorrelation measurement (g$^2$($\tau$)) of the characterised NV centre sites including the corresponding background contributions from the photonic device and the measurement apparatus. Here, the measured photon autocorrelation intercept at zero delay  g$^2$(0) $<$ 0.5 indicates emission dominated by a single photon emitter \cite{berthel2015photophysics, kurtsiefer2000stable}.  We are able to observe single emitter characteristics for both of these reference spots. The thinner substrate has a borderline single photon emission  $g^2(0) = 0.45$ with a higher purity observed for the thicker SiO$_2$ substrate with $g^2(0) = 0.31$.  

It should be noted that in the case of the device with no metal reflector, no discernible g$^2$($\tau$)<1 was observed even for much longer integration times on all Gaussian spots found in Fig.\ref{fig2}, despite the presence of NV centre emission PL spectra and a saturation measurement (Supplementary Information S6). It is thought likely that any bright site was either a multiple-emitter or, for dimmer sites, not enhanced by the metallic mirror, with a larger proportional contribution from the surrounding substrate and measurement apparatus added uncorrelated photons to the signal \cite{neu2011single, beveratos2002room, fu2009observation}.

In summary, in this work, we present the design and process flow of a hybrid materials platform positioning engineered isotopically-enriched ND with photonic devices. In contrast to previous work with electron-beam lithography, we present photolithographically positioning as an accurate and scalable method to combine quantum emitters with other material platforms. By designing and fabricating a broadband metal reflector and tuning the SiO$_2$ spacer thickness, we significantly improved photon collection efficiency, overcoming numerical aperture limitations imposed by cryogenic microscopy. We achieve enhanced signal collection and record $g^2(0)<0.5$ through the hybrid material stack, indicating single photon purity. 

% Finally, we  proposed a hybrid photonic device using silicon nitride (SiN) waveguide in this platform. 

In future, by developing further devices and a set of cryogenic comparative techniques, we can record the NV signal against its background noise as a useful probe of the local environment of the emitter, which will be useful for evaluating the suitability of a range of ND processing, material environments, and photonic structures, in order to improve desired quantum properties of an engineered platform. Future steps will focus on achieving high-resolution line-width measurements to probe photonic coherence, integrating this ND with photonic waveguides \cite{weng2023heterogeneous} for efficient spin-photon coupling, and optimising fabrication conditions to minimise spectral diffusion. This platform represents a step towards quantum technologies based on the solid-state spin-photon interface and could be extrapolated to characterise other ND-hosted emitter platforms such as silicon vacancies that have a higher proportion of light in the ZPL \cite{jantzen2016nanodiamonds} or emerging O-band emitters \cite{mukherjee2023telecom} compatible with telecommunication infrastructure.

\section{Backmatter}
%Backmatter sections should be listed in the order Funding/Acknowledgment/Disclosures/Data Availability Statement/Supplemental Document section. An example of backmatter with each of these sections included is shown below.

\begin{backmatter}
\bmsection{Funding} Engineering and Physical Sciences Research Council (EP/W006685/1, EP/S023607/1).

\bmsection{Acknowledgments} We would like to thank Matthew Markham, Andrew Edmonds, Daniel Twitchen for growing the diamond and Mike Neville, Alex S. Clark, Rowan Hoggarth, Quanzhong Jiang, Pisu Jiang, and Andrew Murray for useful discussions.

\smallskip

\bmsection{Data availability} Data presented in this manuscript is available from the authors on request.  

\bmsection{Disclosures} The authors declare no conflicts of interest.
 
\bibliography{Bib_File/ND_Photonic} % Ensures references are included
\end{backmatter}

\end{document}